\documentclass[amsart]{article}
\usepackage[english]{babel} 
\usepackage[latin1]{inputenc}
\usepackage{orcidlink}
\usepackage{amsmath,amssymb,amsthm,amsfonts,amscd,authblk,dsfont,color,cancel,diagbox,enumerate,epsfig,esint,eucal,fancyhdr,latexsym,lmodern,mathabx,mathrsfs,mathtools,multicol,multirow,musicography,phaistos,skull,simpler-wick,subcaption,tikz-cd,xfrac}
\usepackage[normalem]{ulem}
\DeclareMathOperator{\tr}{tr}

\usepackage[multiuser]{fixme}

\def\bbbone{{\mathchoice {\rm 1\mskip-4mu l} {\rm 1\mskip-4mu l}
		{\rm 1\mskip-4.5mu l} {\rm 1\mskip-5mu l}}}
\def\one{\bbbone}

\usepackage[
    backend=biber,
    style=alphabetic,
    sorting=nyt,
    giveninits=true,
    maxbibnames=99,
    natbib=true
]{biblatex}

\usepackage[all,cmtip]{xy}
\usetikzlibrary{matrix,calc}
\usepackage[autostyle, english = american]{csquotes}
\MakeOuterQuote{"}

\theoremstyle{plain}

\newtheorem{thm}{Theorem}[section]
\newtheorem{lem}[thm]{Lemma}

\theoremstyle{definition}

\newtheorem{notation}[thm]{Notation}

\newtheorem{defn}[thm]{Definition}

\theoremstyle{remark}
\newtheorem{rem}[thm]{Remark}

\title{A Finite-Range, Translation-Invariant Counterexample to Weak Gibbsianity}
\author[a]{Lorenzo Pettinari\thanks{\href{mailto:lorenzo.pettinari@unitn.it}{lorenzo.pettinari@unitn.it}}}

\affil[a]{Dipartimento di Matematica, Universit\`a di Trento and INFN-TIFPA and INdAM, Via Sommarive 14, I-38123 Povo, Italy}

\begin{document}
\maketitle

\begin{abstract}
 We construct a finite-range, translation-invariant counterexample to the Weak Gibbsianity property for quantum spin systems on $\mathbb{Z}^2$. The proof relies on showing how a local potential can induce macroscopically large deviations from an equilibrium configuration after a perturbation is introduced at the boundary of the lattice.
\end{abstract}

\section*{Notation}
We introduce now the basic notation that we will employ in the following.

\begin{notation}[\textbf{Lattice structures}]\label{not: lattice notation}\noindent\begin{itemize}
    \item  We denote by $\Lambda\Subset \mathbb{Z}^2$ finite sublattices of $\mathbb{Z}^2$.
    \item We denote by $X^c$ the complements of $X$ in $\mathbb{Z}^2$: $X^c:= \mathbb{Z}^2\setminus X$.
    \item For any finite set $X\Subset\mathbb{Z}^2$ we denote its cardinality by $|X|\in\mathbb{Z}_+$
\end{itemize}
\end{notation}
\begin{notation}[\textbf{Quasi-local algebraic setting}]
    We denote a net of local algebras indexed by finite subsets of $\mathbb{Z}^2$ by $(\mathcal{A}_X)_{X\Subset\mathbb{Z}^2}$. In this work, for every $i\in\mathbb{Z}^2$ we have $\mathcal{A}_{\{i\}} = M_{2}(\mathbb{C})$ and $$\mathcal{A}_X = \bigotimes_{i\in X}\mathcal{A}_{\{i\}}.$$ The quasi local algebra $\mathcal{A}$ is constructed by standard inductive limit argument from the local algebras $\mathcal{A}_\Lambda$; see \cite[Section 2.6]{BratteliRobinson1987Vol1} and \cite[Section 6.2]{BratteliRobinson1997Vol2} for additional details.
\end{notation}
\begin{notation}[\textbf{Spin matrices and computational basis}] We denote by $S^\alpha_i$, $i\in\mathbb{Z}^2$, $\alpha = \{x,y,z\}$ the spin-$\frac{1}{2}$ angular momentum operators at site $i$\footnote{To avoid confusion we will distinguish between the site label $i$ and the imaginary unit $\mathrm{i}$.}, satisfying the $SU(2)$ commutation relations \begin{equation*}
    [S_i^\alpha,S_j^{\beta}] = \mathrm{i}\delta_{ij}\sum_{\gamma}\epsilon^{\alpha\beta\gamma}S^\gamma_i.
\end{equation*} We denote by $\sigma_i^\alpha = 2S^\alpha_i$ the corresponding Pauli matrices.

    We define the computational basis in a finite region $\Lambda$ as the tensor product over all points of $\Lambda$ of the eigenstates of $\sigma^z_i$. These basis elements will be labeled as $\sigma_\Lambda$, where $\sigma_\Lambda \in \{+,-\}^{\Lambda}$. Alternatively, when we need to specify the configuration we will adopt Dirac's bra-ket notation \begin{equation*}
	\sigma_{\{i\}} = | \pm \rangle_i \qquad i\in \Lambda.
\end{equation*}
\end{notation}
\section{Introduction}
\label{Sec: introduction}

The concept of a \textit{Weak Gibbs state} has its roots in the
thermodynamic formalism of classical statistical mechanics
\cite{Ruelle1969StatisticalMechanics}. This property was subsequently
formulated explicitly and explored in the theory of dynamical systems
\cite{Yuri2000WeakGibbs,Yuri2002WeakGibbs}, as well as in connection
with equilibrium measures and large deviations
\cite{PfisterSullivan2018WeakGibbs,PfisterSullivan2020Asymptotic}.  

Even though they were not precisely formalized before recent times, quantum Weak Gibbs state already appeared in the quantum spin system literature \cite{Araki1976PositiveCone,ArakiIon1974Equivalence}, and more recently found application in the study of \textit{large deviation principles} for quantum lattice spin systems \cite{OgataReyBellet2011}.

More recently, the study of Weak Gibbs states has received renewed
attention in the question of \textit{approach to equilibrium} for quantum statistical mechanics through the works of
Jak\v{s}i\'{c}, Pillet and Tauber
\cite{JaksicPilletTauber2024Approach,JaksicPilletTauber2024Adiabatic}, who gave the first formalized notion for quantum spin systems,
and Jak\v{s}i\'{c}, Pillet and Szczepanek
\cite{JaksicPilletSzczepanek2026Entropy,
JaksicPilletSzczepanek2026Conservation,
JaksicPilletSzczepanek2026ApproachII}.
A standard formulation of Weak Gibbsianity for quantum spin systems
is the following; see \cite[Definition 2.9]{JaksicPilletTauber2024Approach}.

\begin{defn}[\textbf{Weak Gibbs state}]
    Let $\Phi=(\Phi_X)_{X\Subset\mathbb{Z}^d}$
    be a self-adjoint \textit{interaction} for a quantum spin system
    with quasi-local algebra $\mathcal{A}$ and local algebras
    $(\mathcal{A}_X)_{X\Subset\mathbb{Z}^d}$, so that
    $\Phi_X=\Phi_X^*\in\mathcal{A}_X$.

    Fix an inverse temperature $\beta\in\mathbb{R}$. For any finite
    region $\Lambda\Subset\mathbb{Z}^d$, let
    $H_\Lambda:=\sum_{X\subseteq\Lambda}\Phi_X$ and define the
    finite-volume Gibbs state by
   $$
        \omega_\Lambda(A_\Lambda)
        :=
        \frac{\tr[e^{\beta H_\Lambda}A_\Lambda]}
             {\tr[e^{\beta H_\Lambda}]},
        \qquad A_\Lambda\in\mathcal{A}_\Lambda.
$$

    A state $\omega\in\mathcal{S}(\mathcal{A})$ is said to be a
    \textit{Weak Gibbs state} with respect to $\Phi$ at inverse
    temperature $\beta$ if there exists a family of constants
    $(c_\Lambda)_{\Lambda\Subset\mathbb{Z}^d}$, with $c_\Lambda\geq 0$,
    such that
    $$
        \lim_{\Lambda\uparrow\mathbb{Z}^d}
        \frac{c_\Lambda}{|\Lambda|}=0,
    $$
    where the limit is taken along increasing cubes, and
    \begin{equation}\label{eq: intro - Weak gibbsianity}
        e^{-c_\Lambda}\omega_\Lambda
        \leq \omega\restriction_{\mathcal{A}_\Lambda}
        \leq e^{c_\Lambda}\omega_\Lambda
    \end{equation}
    for every finite region $\Lambda\Subset\mathbb{Z}^d$.
    The inequalities are understood in the order on positive
    linear functionals, that is, after evaluation on every
    positive element of $\mathcal{A}_\Lambda$.
\end{defn}
Usually, this definition is given for restricted classes of interactions $\Phi$, such as those that are translation-invariant and exponentially decaying.

It was widely expected, and explicitly conjectured in \cite{JaksicPilletSzczepanek2026ApproachII}, that Weak Gibbsianity should hold for thermal equilibrium states of any quantum spin system on the lattice $\mathbb{Z}^d$, $d\geq 1$, under suitable regularity assumptions on the interaction, such as finite range or exponential decay. Previous results in this direction date back to Araki \cite{Araki1969Gibbs} for one-dimensional lattice systems with finite-range interactions, and to Araki and Ion \cite{ArakiIon1974Equivalence} for arbitrary $d\geq 1$ at sufficiently high temperatures. The Weak Gibbsianity property was explicitly established in these settings in \cite[Theorem 2.11]{JaksicPilletTauber2024Approach}.

In this paper, we show that the Weak Gibbsianity conjecture is false for $d\geq 2$\footnote{The results of this work are formulated for $d=2$, but the extension to arbitrary $d\geq 2$ is trivial.}, even for finite-range, translation-invariant interactions.

The counterexample is constructed starting from the following nearest-neighbors potential
\begin{equation}\label{eq: failure of factorization - potential}
\Phi(X):=J \begin{dcases}
\frac{\one_i-\sigma^z_i}{2}\sigma^x_j
+\sigma^x_i\frac{\one_j-\sigma^z_j}{2}
& \text{if }X=\{i,j\}\text{ and }|i-j|=1,\\
0 & \text{otherwise},
\end{dcases}
\end{equation}
for all $X\Subset\mathbb{Z}^2$. The potential in \eqref{eq: failure of factorization - potential} does not belong to the usual class of XYZ interactions. For every pair of neighboring sites $(i,j)$, it is given by a coupling $J>0$ times the sum of two terms: the tensor product of the projection onto the $|-\rangle_i$ spin state at site $i$, $$P_i^z := \frac{\one_i- \sigma_i^z}{2},$$ and the Pauli matrix $\sigma^x_j$ at site $j$, and the corresponding term with $i$ and $j$ interchanged. Therefore, the action of the interaction term $P_i^z\sigma_j^x$ on the computational basis is as follows \begin{equation}\label{eq: failure of factorization - coefficients}
P_i^z\sigma_j^x |\sigma_i,\sigma_j\rangle = (\delta_{\sigma_i,-} )|\sigma_i,-\sigma_j\rangle.
\end{equation} We see that every interaction potential has positive matrix coefficients in the computational basis. 

For every finite subset $\Lambda\Subset\mathbb{Z}^2$, we denote the Hamiltonian of the model as \begin{align*}
H_\Lambda = \sum_{X\subseteq \Lambda}\Phi(X) =J \sum_{\substack{i,j\in \Lambda\\ |i-j|=1}} P_i^z \sigma_j^x 
\end{align*} 

By standard results \cite[Section 6.2]{BratteliRobinson1997Vol2}, the net of Gibbs states $(\omega_\Lambda)_{\Lambda\Subset\mathbb{Z}^2}$ for the interaction in \eqref{eq: failure of factorization - potential} admits convergent subsequences to $\beta$-KMS states on $\mathcal{A}$ for the infinite volume dynamics generated by $\Phi$.

The central result of the paper is formulated in the following theorem. 
\begin{thm}[\textbf{Counterexample to the Weak Gibbsianity property}]\label{prop: failure of factorization - upper bound}
Let $\omega$ be any limit point of the net of finite volume Gibbs states $(\omega_\Lambda)_{\Lambda\Subset\mathbb{Z}^2}$ for the interaction potential in \eqref{eq: failure of factorization - potential}. Suppose that the inverse temperature $\beta$ satisfies \begin{equation*}
J\beta > 2^{\frac{11}{2}}6^{\frac{1}{4}}.
\end{equation*} Then, there exists a positive constant $C_{J\beta}>0$ and an  increasing sequence of squares $(X_k)_{k\in\mathbb{N}}$ converging to $\mathbb{Z}^2$ such that \begin{equation}\label{eq: intro - failure of Weak gibbs}
\omega(\one_{\{\frac{1}{2}\}}(m_{X_k}^z)) \geq e^{C_{J\beta}|X_k|}\omega_{X_k}(\one_{\{\frac{1}{2}\}}(m_{X_k}^z)),
\end{equation} where $$m^z_{X_k} = \frac{1}{|X_k|}\sum_{i\in X_k}S^z_i$$ is the mean magnetization along the $z$-axis in the region $X_k$.
\end{thm} 

We see that Equation \eqref{eq: intro - failure of Weak gibbs} provides an explicit counterexample to the upper bound in the Weak Gibbsianity condition \eqref{eq: intro - Weak gibbsianity} for any limit point of the net of Gibbs states $(\omega_\Lambda)_{\Lambda\Subset\mathbb{Z}^2}$.

The proof of the counterexample in Theorem \ref{prop: failure of factorization - upper bound} relies on techniques similar to those used to establish the blow-up of complex-time evolution in lattice spin systems in dimensions $d\geq 2$. These techniques were originally developed by Bouch \cite{Bouch2015} for a specific quantum spin model. Tasaki \cite{Tasaki2024Bouch} subsequently provided a clearer exposition of Bouch's construction and, in a later revision, incorporated an LLM-generated proof extending the combinatorial result from a particular sequence of tree sizes to every positive integer size. It is possible that the "Extraction lemma" \cite[Lemma 2.2]{Tasaki2024Bouch} could be used to improve our proof, but we did not check this. Shiraishi and Tasaki \cite[Appendix A.4, Theorem A.6]{ShiraishiTasaki2026} also employed Bouch's construction to extend the complex-time blow-up result to a general class of XY and XYZ models.

In Section \ref{Sec: proof of the counterexample}, we prove Theorem \ref{prop: failure of factorization - upper bound} and discuss some of its immediate consequences.
\section{Proof of the counterexample}\label{Sec: proof of the counterexample}

 In this section, we will present a proof of Theorem \ref{prop: failure of factorization - upper bound}, showing that there exist  finite-range, translation-invariant models for which the Weak Gibbsianity property fails on local observables. We will work in a two dimensional lattice $\mathbb{Z}^2$ by first restricting to  a finite subregion of the lattice $\Lambda\Subset\mathbb{Z}^2$. It is fairly easy to extend all the results to  $d\geq2$ dimensional lattices.

We need a few preliminary observations. The first one concerns the mean magnetization along the third direction, and its relation with the interaction potential defined in Equation \ref{eq: failure of factorization - potential}.
\begin{rem} 
	We can immediately read from Equation \ref{eq: failure of factorization - coefficients} that the "all spin up" state $$\Omega_\Lambda^+:= \bigotimes_{i\in\Lambda}\left| +\right\rangle_i$$ is in the Kernel of $H_\Lambda$\begin{equation*}
		H_\Lambda \Omega_\Lambda^+ = 0.
	\end{equation*} 
	The projection onto $\Omega_\Lambda^+$ is also the projection corresponding to the value $\frac{1}{2}$ of the mean magnetization along the third direction $$m_\Lambda^z = \frac{1}{|\Lambda|}\sum_{i\in\Lambda}S_i^z,\qquad  |\Omega_\Lambda^+\rangle\langle\Omega_\Lambda^+| = \one_{\{\frac{1}{2}\}}(m_\Lambda^z).$$
\end{rem}

Next, we fix a positive inverse temperature $\beta>0$ and we consider the finite volume Gibbs state \begin{equation*}
\omega_{\Lambda}(A_\Lambda) := \frac{\tr[e^{\beta H_\Lambda}A_\Lambda]}{\tr[e^{\beta H_\Lambda}]}
\end{equation*} for any region $\Lambda\Subset\mathbb{Z}^2$. \begin{rem}\label{obs: failure of factorization - 1}
For any $X\Subset\mathbb{Z}^2$ we have \begin{equation*}
	\omega_X(\one_{\{\frac{1}{2}\}}(m_X^z)) = \frac{1}{\tr[e^{\beta H_X}]} = Z_X^{-1},
\end{equation*} where $Z_X$ is the partition function for the model in Equation \ref{eq: failure of factorization - potential}, localized in $X\Subset\mathbb{Z}^2$.
\end{rem} 

We consider the boundary interaction for any finite subregion $X\subset\Lambda$ \begin{equation*}
	W^\Lambda_X := \sum_{\substack{Y\subseteq\Lambda, Y\cap X\neq \emptyset\\ Y\cap X^c\neq \emptyset}} \Phi(Y) = J\sum_{\substack{i \in X^c, j\in X\\ |i-j|=1}}P_i^z\sigma^x_j + \sigma_i^xP_j^z.
\end{equation*} We can bound the norm of $W^\Lambda_X$ as \begin{equation*}
\|W^\Lambda_X\| \leq 2J|\partial X|,
\end{equation*} where $$\partial X:= \{\{i,j\}\subset\mathbb{Z}^2 \mid |i-j| = 1,\, i\in X,\, j\in X^c\}$$ is the boundary of $X$. 
\begin{rem}\label{obs: failure of factorization - 2}
	Since the interaction potential $\Phi$ has positive matrix elements with respect to the computational basis, it easily follows that \begin{equation*}
	\tr[e^{\beta H_\Lambda } \one_{\{\frac{1}{2}\}}(m_X^z)]\geq \tr[e^{\beta H_{\Lambda\setminus Y}}] \tr[e^{\beta H_{Y}}\one_{\{\frac{1}{2}\}}(m_X^z)]
	\end{equation*} for any $X\subseteq Y\subseteq \Lambda$. Moreover, by a standard Golden-Thompson inequality, the partition function satisfies \begin{equation*}
	Z_\Lambda \leq e^{2J\beta|\partial Y|} Z_Y Z_{\Lambda\setminus Y}.
	\end{equation*} Hence, we see that \begin{equation*}
	\omega_\Lambda(\one_{\{\frac{1}{2}\}}(m_X^z)) \geq e^{-2J\beta|\partial Y|}\omega_Y(\one_{\{\frac{1}{2}\}}(m_X^z)).
	\end{equation*}
\end{rem} 

We can now outline the main steps in the construction of our counterexample. Let us define for any finite set $X\subset \Lambda$  $$X(i):= X\cup \{i\},\qquad i\in \Lambda,$$ where $i\notin X$ and we suppose that there exists a unique $j\in X$ such that $|i-j| = 1$. Combining Observations \ref{obs: failure of factorization - 1} and \ref{obs: failure of factorization - 2} we arrive at the inequality \begin{align*}
	\omega_\Lambda(\one_{\{\frac{1}{2}\}}(m_X^z)) \geq& e^{-2\beta J |\partial X(i)|}\tr[e^{\beta H_{X(i)}}\one_{\{\frac{1}{2}\} (m_X^z)}] Z_{X(i)}^{-1} \\ 
	\geq& \frac{1}{2}e^{-2J\beta |\partial X(i)|-2J\beta}\tr[e^{\beta H_{X(i)}}\one_{\{\frac{1}{2}\} (m_X^z)}] \omega_X(\one_{\{\frac{1}{2}\}}(m^z_X)).
\end{align*} To further estimate the last line, let us choose an element of the computational basis  $\sigma_{X(i)}$ in the range of the projector $\one_{\{\frac{1}{2}\}}(m^z_X)$, i.e. $\one_{\{\frac{1}{2}\}}(m^z_X)\sigma_{X(i)}=\sigma_{X(i)}$. This essentially means that $\sigma_{X(i)}$ coincide with $\Omega_X^+$ on $X$, while it can be chosen arbitrarily on $X(i)\setminus X$. Then, for any other computational basis element $\xi_{X(i)}$  we have\begin{align*}
\tr[e^{\beta H_{X(i)} } \one_{\{\frac{1}{2}\}}] \geq& \left \langle \sigma_{X(i)} \mid  e^{\beta H_{X(i)}}\sigma_{X(i)}\right\rangle\\
\geq &\left|  \left \langle \sigma_{X(i)} \mid  e^{\frac{\beta}{2} H_{X(i)}}\xi_{X(i)}\right\rangle\right|^2.
\end{align*} Using again that the matrix elements of $H_\Lambda$ are positive in the computational basis we can write \begin{align*}
\left| \left \langle \sigma_{X(i)} \mid  e^{\frac{\beta}{2} H_{X(i)}}\xi_{X(i)}\right\rangle\right| =& \left \langle \sigma_{X(i)} \mid  e^{\frac{\beta}{2} H_{X(i)}}\xi_{X(i)}\right\rangle \\
\geq & \frac{1}{n!}\left \langle \sigma_{X(i)} \mid  \left(\frac{\beta}{2}H_{X(i)}\right)^n \xi_{X(i)}\right\rangle,
\end{align*} where, in the last line, we have estimated the full Taylor series of the exponential using its term of order $n$, for arbitrary $n\geq 0$.  The last bracket can be written as a sum of terms of the form \begin{equation*}
\frac{1}{n!}\left(\frac{\beta J}{2}\right)^n \left\langle \sigma_{X(i)} \mid   P^z_{l_n}\sigma^x_{k_n} \dots P^z_{l_1}\sigma^x_{k_1} \xi_{X(i)}\right\rangle,\qquad |l_m-k_m|=1 ,\, m = 1,\dots, n.
\end{equation*} We can interpret the above as a path $(k_1,l_1\dots k_n,l_n)$  on the set $X(i)$, where at every step $m = 1,\dots, n$ the spin at a site $k_m$ is reversed if the spin at site $l_m$ is $|-\rangle_{l_m}$. The only non zero contribution will be those arising from paths changing the configuration $\xi_{X(i)}$ into the configuration $\sigma_{X(i)}$. For all such paths, the bracket will be equal to one. Therefore, it remains to estimate the number of these admissible paths. If this number, denoted by $N(\xi_{X(i)},\sigma_{X(i)},n)$, were to satisfy
\begin{equation*}
N(\xi_{X(i)},\sigma_{X(i)},|X|) \geq \frac{|X|!}{C^{|X|}},
\end{equation*}
for some constant $C>0$, then we would obtain the estimate\begin{align*}
\tr[e^{\beta H_{X(i)}}\one_{\{\frac{1}{2}\}}(m_X^z)] \geq \left(\frac{\beta J}{2}\right)^{2|X|}\left(\frac{N(\xi_{X(i)},\sigma_{X(i)},|X|)}{|X|!}\right)^2 \geq  \left(\frac{\beta J}{2C}\right)^{2|X|}.
\end{align*} Putting all together, this would imply the inequality \begin{equation}\label{eq: failure of factorization - final estimate}
\omega_\Lambda(\one_{\{\frac{1}{2}\}}(m_X^z)) \geq e^{2|X|\ln\left(\frac{\beta J}{2 C}\right)} e^{-2J\beta|\partial X(i)|-2J\beta-\ln(2)} \omega_X(\one_{\{\frac{1}{2}\}}(m_X^z)).
\end{equation}
We will now describe in detail the estimate for the number $N(\xi_{X(i)},\sigma_{X(i)},|X|)$. The core idea of our construction is the following. To allow for many different paths connecting the configurations $\sigma_{X(i)}$ and $\xi_{X(i)}$, these vectors must differ  at a macroscopic number of points in the lattice. The second requirement is that there should be many distinct ways to change from one configuration to the other by repeatedly applying the potential $P^z_i \sigma^x_j$. As we will see, these conditions are satisfied by our choice of the potential \eqref{eq: failure of factorization - potential}. Indeed, the path $P^z_{l_n}\sigma^x_{k_n} \dots P^z_{l_1}\sigma^x_{k_1} \xi_{X(i)}$ can have many different branches, giving rise to a fractal tree structure. This procedure follows the idea in Bouch's analysis of complex time singularity \cite{Bouch2015}, although the details of the computations and the choice of the tree differ from those of Bouch.
\subsection{Construction of the spanning tree}
We consider a sequence of lattice squares $X_k$ constructed as follows. Fix $L_k = 2^k$ for $k\in\mathbb{N}$ and take $X_k = \{0,\dots L_k-1\}^2$. We enlarge it by adding the point $(-1,0)$, so that its enlarged version is given by $X_k(-1,0) = \{0,\dots , L_k-1\}^2\cup\{(-1,0)\}$. The box's sides length satisfies $2L_{k-1} = L_k$. Our choice for the vectors $\xi_{X_k(-1,0)}$ and $\sigma_{X_k(-1,0)}$ is the following \begin{align*}
\sigma_{X_k(-1,0)} =& |-\rangle_{(-1,0)}\bigotimes_{i\in X_k }| +\rangle_i,\\ 
\xi_{X_k(-1,0)}=& \bigotimes_{i\in X_k(-1,0)}|-\rangle_i.
\end{align*}
We see that the two configurations overlap only at the vertex $(-1,0)$.  This is the starting point of our construction. Iteratively applying the operators $P_{(-1,0)}^z\sigma^x_{(0,0)}\to P_{(0,0)}^z\sigma^x_{(1,0)}\to \dots $ we will construct  a path of operators covering the whole square $X_k$ and flipping the spin at every point by repeated actions \begin{align*}
&P_{(-1,0)}^z\sigma^x_{(0,0)} |-\rangle_{(-1,0)}\bigotimes_{i\in X_k }| +\rangle_i \\ 
=& |-\rangle_{(-1,0)}\otimes |-\rangle_{(0,0)}\bigotimes_{i\in X_k\setminus\{(0,0)\} }| +\rangle_i 
\end{align*} and so on. This path must contain many branches, otherwise the number of possible combinations that generate the admissible paths would note be sufficient to violate the Weak Gibbsianity property. We will rely on a fractal tree construction, inspired by the results presented in \cite{TAPP2026104325}\footnote{We remark that different spanning tree constructions work in building our counterexample. The one described here is the simplest one we could find in our trials. It also seems to lead to an optimal bound on the inverse temperature, even though we do not have a rigorous proof of this.}. The resulting path is constructed by inductively building  a spanning tree $T_k = (V_k, E_k)$. here $V_k = X_k$ are the vertices of the tree, whereas $E_k$ are the edges. 

Let us describe the first iterative step of the construction. We take $T_1$ to be the spanning tree for $X_1$ with edges $$\{e_{(0,0)-(1,0)},\,e_{(1,0)-(1,1)},\, e_{(0,1)-(1,1)}\} = E_1. $$ To construct $T_2$, we attach four identical copies of $T_1$ at the four corners of $X_2$, and we connect these by another copy of $T_1$ in the middle square.  We repeat this procedure iteratively. To construct the spanning tree $T_k$ in $X_k$, we take four translated copies of $T_{k-1}$ and we attach these to the four corners of $X_k$. We connect these four trees thorough another copy of $T_1$ in the middle of the square $X_k$. The first tree steps of the construction are illustrated in Figure \ref{fig: failure of factorization - fractal graph}. Note that the resulting graph is always connected and contains all vertices of $X_k$.

\begin{figure}\centering
\begin{tikzpicture}[scale=0.8,
    transform shape,
  line width=1.15pt,
  line cap=round,
  line join=round,
  every node/.style={font=\normalsize,inner sep=0pt},
  vertex/.style={circle,fill=treeblack,minimum size=2.7pt,inner sep=0pt}
]
\definecolor{treeblack}{RGB}{34,31,32}
\definecolor{treered}{RGB}{237,28,36}
\definecolor{treecyan}{RGB}{0,174,239}

\begin{scope}[x=4.2cm,y=4.2cm]
  \node[anchor=south] at (0.5,1.1) {$k=1\;\cdot\;2\times2$};
  \draw[treeblack] (0,1) -- (1,1) -- (1,0) -- (0,0);
  \path
    (0,1) node[vertex]{} (1,1) node[vertex]{}
    (0,0) node[vertex]{} (1,0) node[vertex]{};
\end{scope}

\begin{scope}[xshift=5.4cm,x=1.4cm,y=1.4cm]
  \node[anchor=south] at (1.5,3.3) {$k=2\;\cdot\;4\times4$};
  \draw[treeblack] (0,3) -- (1,3) -- (1,2) -- (0,2);
  \draw[treeblack] (2,3) -- (3,3) -- (3,2) -- (2,2);
  \draw[treeblack] (0,1) -- (1,1) -- (1,0) -- (0,0);
  \draw[treeblack] (2,1) -- (3,1) -- (3,0) -- (2,0);
  \draw[treered] (1,2) -- (2,2) -- (2,1) -- (1,1);
  \path
    (0,3) node[vertex]{} (1,3) node[vertex]{} (2,3) node[vertex]{} (3,3) node[vertex]{}
    (0,2) node[vertex]{} (1,2) node[vertex]{} (2,2) node[vertex]{} (3,2) node[vertex]{}
    (0,1) node[vertex]{} (1,1) node[vertex]{} (2,1) node[vertex]{} (3,1) node[vertex]{}
    (0,0) node[vertex]{} (1,0) node[vertex]{} (2,0) node[vertex]{} (3,0) node[vertex]{};
\end{scope}

\begin{scope}[xshift=10.8cm,x=0.6cm,y=0.6cm]
  \node[anchor=south] at (3.5,7.7) {$k=3\;\cdot\;8\times8$};
  \draw[treeblack] (0,7) -- (1,7) -- (1,6) -- (0,6);
  \draw[treeblack] (2,7) -- (3,7) -- (3,6) -- (2,6);
  \draw[treeblack] (4,7) -- (5,7) -- (5,6) -- (4,6);
  \draw[treeblack] (6,7) -- (7,7) -- (7,6) -- (6,6);
  \draw[treeblack] (0,5) -- (1,5) -- (1,4) -- (0,4);
  \draw[treeblack] (2,5) -- (3,5) -- (3,4) -- (2,4);
  \draw[treeblack] (4,5) -- (5,5) -- (5,4) -- (4,4);
  \draw[treeblack] (6,5) -- (7,5) -- (7,4) -- (6,4);
  \draw[treeblack] (0,3) -- (1,3) -- (1,2) -- (0,2);
  \draw[treeblack] (2,3) -- (3,3) -- (3,2) -- (2,2);
  \draw[treeblack] (4,3) -- (5,3) -- (5,2) -- (4,2);
  \draw[treeblack] (6,3) -- (7,3) -- (7,2) -- (6,2);
  \draw[treeblack] (0,1) -- (1,1) -- (1,0) -- (0,0);
  \draw[treeblack] (2,1) -- (3,1) -- (3,0) -- (2,0);
  \draw[treeblack] (4,1) -- (5,1) -- (5,0) -- (4,0);
  \draw[treeblack] (6,1) -- (7,1) -- (7,0) -- (6,0);
  \draw[treered] (1,6) -- (2,6) -- (2,5) -- (1,5);
  \draw[treered] (5,6) -- (6,6) -- (6,5) -- (5,5);
  \draw[treered] (1,2) -- (2,2) -- (2,1) -- (1,1);
  \draw[treered] (5,2) -- (6,2) -- (6,1) -- (5,1);
  \draw[treecyan] (3,4) -- (4,4) -- (4,3) -- (3,3);
  \path
    (0,7) node[vertex]{} (1,7) node[vertex]{} (2,7) node[vertex]{} (3,7) node[vertex]{} (4,7) node[vertex]{} (5,7) node[vertex]{} (6,7) node[vertex]{} (7,7) node[vertex]{}
    (0,6) node[vertex]{} (1,6) node[vertex]{} (2,6) node[vertex]{} (3,6) node[vertex]{} (4,6) node[vertex]{} (5,6) node[vertex]{} (6,6) node[vertex]{} (7,6) node[vertex]{}
    (0,5) node[vertex]{} (1,5) node[vertex]{} (2,5) node[vertex]{} (3,5) node[vertex]{} (4,5) node[vertex]{} (5,5) node[vertex]{} (6,5) node[vertex]{} (7,5) node[vertex]{}
    (0,4) node[vertex]{} (1,4) node[vertex]{} (2,4) node[vertex]{} (3,4) node[vertex]{} (4,4) node[vertex]{} (5,4) node[vertex]{} (6,4) node[vertex]{} (7,4) node[vertex]{}
    (0,3) node[vertex]{} (1,3) node[vertex]{} (2,3) node[vertex]{} (3,3) node[vertex]{} (4,3) node[vertex]{} (5,3) node[vertex]{} (6,3) node[vertex]{} (7,3) node[vertex]{}
    (0,2) node[vertex]{} (1,2) node[vertex]{} (2,2) node[vertex]{} (3,2) node[vertex]{} (4,2) node[vertex]{} (5,2) node[vertex]{} (6,2) node[vertex]{} (7,2) node[vertex]{}
    (0,1) node[vertex]{} (1,1) node[vertex]{} (2,1) node[vertex]{} (3,1) node[vertex]{} (4,1) node[vertex]{} (5,1) node[vertex]{} (6,1) node[vertex]{} (7,1) node[vertex]{}
    (0,0) node[vertex]{} (1,0) node[vertex]{} (2,0) node[vertex]{} (3,0) node[vertex]{} (4,0) node[vertex]{} (5,0) node[vertex]{} (6,0) node[vertex]{} (7,0) node[vertex]{};
\end{scope}
\end{tikzpicture}
\caption{Fractal spanning tree construction. This figure is a partial reproduction of \cite[Figure 1]{TAPP2026104325}. }
\label{fig: failure of factorization - fractal graph}
\end{figure}

Let us prove a first geometric lemma that will be needed for future estimates \begin{lem}\label{lemma: failure of factorization - Dk}
	Let $T_k$ be the spanning tree defined above. Call $D_k$ the diameter of $T_k$ defined as the longest non-intersecting path along the edges of $T_k$. Then,  \begin{equation*}
		D_k \leq 3(2^{k}-1).
	\end{equation*}
    \end{lem}
    \begin{proof}It is clear that $D_1 = 3$. For any two points $x,y\in X_k$, these satisfy one of the following: \begin{enumerate}
	\item Both $x$ and $y$ belong to the central square $T_1$;
	\item One of the two points belongs to the central square say $x \in T_1$, while the other belongs to one of the four copies of $T_{k-1}$;
	\item They both belong to the same copy of $T_{k-1}$;
	\item They belong to different copies of $T_{k-1}$.
	\end{enumerate}In the first case, their distance along the edges of $T_k$ is at most $3$. As for the second case, their distance is less than the number of edges that one of the two points has to travel to reach the central square, plus the diameter of the  square. This is less than $D_{k-1}+3$. In the third case , their distance is always less than $D_{k-1}$. In the fourth case, their distance is always less than the number of edges it takes for both of them to travel to the cross, plus the length of the cross. This is less than $2D_{k-1}+3$. Therefore, we see that \begin{equation*}
	D_k \leq 2D_{k-1}+ 3.
	\end{equation*}Iterating this inequality $k-1$ times we get \begin{equation*}
	D_k \leq  3\sum_{n=0}^{k-1}2^n = 3(2^k-1).
	\end{equation*} This concludes the proof of the Lemma.\end{proof}

Next, we count the number of ways to construct the tree $T_k$ as a nested tree, starting from a given root $r\in X_k$. In \cite[Lemma 6.2]{Bouch2015}, Bouch proved that the number of different constructions leading to the same rooted tree with root $r$ and edges $e_1^r,\dots,e_n^r$ is given by
\begin{equation*}
	\frac{n!}{\prod_{i=1}^n w(e_i^r)},
\end{equation*}
where $w(e_i^r)$ is the weight of the edge $e_i^r$, defined as one plus the number of its descendants. These are the edges that follow $e_i^r$ as we move away from the root $r$. Therefore, after relabeling the edges of $T_k$ as $e_1^r,\dots,e_{|X_k|-1}^r$, we obtain a lower bound on the number of possible constructions by finding an upper bound on
\begin{equation*}
	\prod_{i=1}^{|X_k|-1} w(e_i^r).
\end{equation*}
Let us define
\begin{equation*}
	W_k := \sup_{r\in T_k} \sum_{i=1}^{|X_k|-1}\ln(w(e_i^r)).
\end{equation*}
We will estimate $W_k$ iteratively. In particular, we shall prove \begin{lem}\label{lemma: failure of factorization - Fk}
Take $W_k$ as above. Then, \begin{equation*}
	W_k \leq \left(\frac{9}{2}\ln(2)+\frac{1}{4}\ln(6)\right)|X_k|
\end{equation*}
\end{lem}
\begin{proof}
	We distinguish two cases. 
	\noindent
	$i)$ We place the root of $T_k$ on the central square $T_1$. Then, every edges on the central tree will have necessarily less than $|X_k|$ descendant (this is the trivial bound). The product of the weights over all the edges of the four copies $T_{k-1}$ is bounded by $\exp{W_{k-1}}$ by definition. Therefore, summing all the contributions we have \begin{equation*}
		\sum_{i=1}^{|X_k|-1}\ln(w(e_i^r)) \leq 4W_{k-1} + 3\ln(|X_k|)
	\end{equation*}
	\noindent
	$ii)$ Suppose we put the root on one of the four squares $T_{k-1}$. The unique path connecting the root with the central copy of $T_1$ contains less than $D_{k-1}$ edges. Each of these will have less than $|X_k|$ descendants. All the other branches starting from this path give a total contribution less than $\exp{W_{k-1}}$. We can repeat the same argument as in $i)$ on the edges of the central square $T_1$ and of the other three copies of $T_{k-1}$. Therefore,  \begin{align*}
			\sum_{i=1}^{|X_k|-1}\ln(w(e_i^r)) \leq &  4W_{k-1}  +(3+ D_{k-1})\ln(|X_k|) \\ 
            \leq & 4W_{k-1} + 3\ln(2)2^{k}k
	\end{align*} where we have used Lemma \ref{lemma: failure of factorization - Dk} to estimate $D_{k-1}$. Since $W_1 = \ln (6)$, iterating the above inequality we get \begin{align*}
	W_k\leq & 3\ln(2)2^{2k}\sum_{n=2}^k \frac{n}{2^n} +2^{2k-2} \ln( 6)\\ 
	 \leq &3\ln(2)2^{2k}\sum_{n=2}^{+\infty}\frac{n}{2^n} +2^{2k-2} \ln (6)= \left(\frac{9}{2}\ln(2)+ \frac{1}{4}\ln(6)\right)|X_k|.
	\end{align*}This last estimate concludes the proof.
\end{proof} 
\subsection{Upper bound on the Gibbs state}
As we already argued in the previous subsection, if we start with a configuration $\sigma_{X_k(-1,0)}$ with all spin up, except for the spin at $(-1,0)$, the contribution to $$\left\langle \xi_{X_k(-1,0)} \mid H_{X_k(-1,0)}^{|X_{k}|}\sigma_{X_k(-1,0)} \right\rangle$$ consists in all the possible paths of operators $P_i^z\sigma^x_j$, $|i-j|=1$ flipping all the spin in $X_k$. Some of the admissible choices are given by the all the possible paths generating the rooted tree $T_k$ with root in $(-1,0)$; see Figure \ref{fig: failure of factorization - rooted-tree} 

\begin{figure}[htbp]
\centering
\begin{tikzpicture}[
    x=0.65cm,y=0.65cm,
    line width=1.1pt,
    line cap=round,
    line join=round,
    vertex/.style={
        circle,fill=treeblack,
        inner sep=0pt,minimum size=2.7pt
    },
    ambient/.style={
        circle,fill=gray!55,
        inner sep=0pt,minimum size=2.2pt
    }
]
\definecolor{treeblack}{RGB}{34,31,32}
\definecolor{treegold}{RGB}{212,160,23}

\draw[draw=gray!55,fill=gray!3,line width=0.5pt]
    (-1.45,-1.45) rectangle (8.45,8.45);

\node[anchor=south] at (3.5,8.7)
    {$k=3\qquad X_k(-1,0)$};
\node at (3.5,7.55) {$T_k$};

\path
    (-1,8) node[ambient]{} (0,8) node[ambient]{}
    (1,8) node[ambient]{} (2,8) node[ambient]{}
    (3,8) node[ambient]{} (4,8) node[ambient]{}
    (5,8) node[ambient]{} (6,8) node[ambient]{}
    (7,8) node[ambient]{} (8,8) node[ambient]{}

    (-1,-1) node[ambient]{} (0,-1) node[ambient]{}
    (1,-1) node[ambient]{} (2,-1) node[ambient]{}
    (3,-1) node[ambient]{} (4,-1) node[ambient]{}
    (5,-1) node[ambient]{} (6,-1) node[ambient]{}
    (7,-1) node[ambient]{} (8,-1) node[ambient]{}

    (-1,1) node[ambient]{} (-1,2) node[ambient]{}
    (-1,3) node[ambient]{} (-1,4) node[ambient]{}
    (-1,5) node[ambient]{} (-1,6) node[ambient]{}
    (-1,7) node[ambient]{}

    (8,0) node[ambient]{} (8,1) node[ambient]{}
    (8,2) node[ambient]{} (8,3) node[ambient]{}
    (8,4) node[ambient]{} (8,5) node[ambient]{}
    (8,6) node[ambient]{} (8,7) node[ambient]{};

\draw[treeblack] (0,7) -- (1,7) -- (1,6) -- (0,6);
\draw[treeblack] (2,7) -- (3,7) -- (3,6) -- (2,6);
\draw[treeblack] (4,7) -- (5,7) -- (5,6) -- (4,6);
\draw[treeblack] (6,7) -- (7,7) -- (7,6) -- (6,6);

\draw[treeblack] (0,5) -- (1,5) -- (1,4) -- (0,4);
\draw[treeblack] (2,5) -- (3,5) -- (3,4) -- (2,4);
\draw[treeblack] (4,5) -- (5,5) -- (5,4) -- (4,4);
\draw[treeblack] (6,5) -- (7,5) -- (7,4) -- (6,4);

\draw[treeblack] (0,3) -- (1,3) -- (1,2) -- (0,2);
\draw[treeblack] (2,3) -- (3,3) -- (3,2) -- (2,2);
\draw[treeblack] (4,3) -- (5,3) -- (5,2) -- (4,2);
\draw[treeblack] (6,3) -- (7,3) -- (7,2) -- (6,2);

\draw[treeblack] (0,1) -- (1,1) -- (1,0) -- (0,0);
\draw[treeblack] (2,1) -- (3,1) -- (3,0) -- (2,0);
\draw[treeblack] (4,1) -- (5,1) -- (5,0) -- (4,0);
\draw[treeblack] (6,1) -- (7,1) -- (7,0) -- (6,0);

\draw[treeblack] (1,6) -- (2,6) -- (2,5) -- (1,5);
\draw[treeblack]  (5,6) -- (6,6) -- (6,5) -- (5,5);
\draw[treeblack]  (1,2) -- (2,2) -- (2,1) -- (1,1);
\draw[treeblack]  (5,2) -- (6,2) -- (6,1) -- (5,1);

\draw[treeblack]  (3,4) -- (4,4) -- (4,3) -- (3,3);

\draw[treegold,line width=1.5pt,->] (-1,0) -- (0,0);

\path
    (0,7) node[vertex]{} (1,7) node[vertex]{}
    (2,7) node[vertex]{} (3,7) node[vertex]{}
    (4,7) node[vertex]{} (5,7) node[vertex]{}
    (6,7) node[vertex]{} (7,7) node[vertex]{}

    (0,6) node[vertex]{} (1,6) node[vertex]{}
    (2,6) node[vertex]{} (3,6) node[vertex]{}
    (4,6) node[vertex]{} (5,6) node[vertex]{}
    (6,6) node[vertex]{} (7,6) node[vertex]{}

    (0,5) node[vertex]{} (1,5) node[vertex]{}
    (2,5) node[vertex]{} (3,5) node[vertex]{}
    (4,5) node[vertex]{} (5,5) node[vertex]{}
    (6,5) node[vertex]{} (7,5) node[vertex]{}

    (0,4) node[vertex]{} (1,4) node[vertex]{}
    (2,4) node[vertex]{} (3,4) node[vertex]{}
    (4,4) node[vertex]{} (5,4) node[vertex]{}
    (6,4) node[vertex]{} (7,4) node[vertex]{}

    (0,3) node[vertex]{} (1,3) node[vertex]{}
    (2,3) node[vertex]{} (3,3) node[vertex]{}
    (4,3) node[vertex]{} (5,3) node[vertex]{}
    (6,3) node[vertex]{} (7,3) node[vertex]{}

    (0,2) node[vertex]{} (1,2) node[vertex]{}
    (2,2) node[vertex]{} (3,2) node[vertex]{}
    (4,2) node[vertex]{} (5,2) node[vertex]{}
    (6,2) node[vertex]{} (7,2) node[vertex]{}

    (0,1) node[vertex]{} (1,1) node[vertex]{}
    (2,1) node[vertex]{} (3,1) node[vertex]{}
    (4,1) node[vertex]{} (5,1) node[vertex]{}
    (6,1) node[vertex]{} (7,1) node[vertex]{}

    (0,0) node[vertex]{} (1,0) node[vertex]{}
    (2,0) node[vertex]{} (3,0) node[vertex]{}
    (4,0) node[vertex]{} (5,0) node[vertex]{}
    (6,0) node[vertex]{} (7,0) node[vertex]{};

\node at (-1,-.6) {$-$};
\node at (0,-.6) {$+$};

\node[vertex,minimum size=3.3pt] at (-1,0) {};
\node[anchor=east,inner sep=0pt] at (-1.65,0) {$r$};

\end{tikzpicture}
\caption{Path construction.}
\label{fig: failure of factorization - rooted-tree}
\end{figure}

Combining Lemma \ref{lemma: failure of factorization - Fk} with Bouch's  argument, we see that the number of way to construct $T_k$ satisfies \begin{equation*}
N(\xi_{X_k(-1,0)},\sigma_{X_k(-1,0)}, |X_k|) \geq |X_k|!|X_k|^{-1}  e^{-\frac{9}{2}\ln(2)|X_k|-\frac{1}{4}\ln(6)|X_k|} .
\end{equation*} Finally, inserting the latter estimate in  \eqref{eq: failure of factorization - final estimate} we conclude \begin{equation*}
\omega_\Lambda(\one_{\{\frac{1}{2}\}}(m_{X_k}^z)) \geq e^{2|X_k|\ln\left(\frac{\beta J}{2^{\frac{11}{2}}6^{\frac{1}{4}}}\right) -2J\beta|\partial X_k(-1,0)| - 2\ln(|X_k|) -\ln(2)-2J\beta}\omega_{X_k}(\one_{\{\frac{1}{2}\}}(m_{X_k}^z)).
\end{equation*} Since this inequality holds for any $\Lambda\supseteq X_k(-1,0)$, we can take a converging subsequence for  $\omega_\Lambda$ as $\Lambda\uparrow\mathbb{Z}^2$. To recover the exact statement of Theorem \ref{prop: failure of factorization - upper bound} it sufficient to repeat the construction using translated  boxes $X_k$ that eventually cover the whole lattice $\mathbb{Z}^2$. 

Thus, we have proved Theorem \ref{prop: failure of factorization - upper bound}.
\begin{rem}[\textbf{Invariance under rotations}]
  Since the magnetization commutes with any rotation around the $z$-axis, the same result holds for all the rotated Hamiltonian \begin{equation*}
     \prod_{i\in\Lambda} U_{\theta,i} H_\Lambda \prod_{i\in\Lambda}U_{\theta,i}^*, \qquad U_{\theta,i} = e^{-\mathrm{i}\theta_i S^z_i}.
  \end{equation*} In particular, by choosing $\theta_i = \pi$ for every $i\in\Lambda$ we can change the sign of the coupling $J$ (or equivalently, of the inverse temperature $\beta$). In a similar fashion, the same result holds for the model where we substitute all the projections $P^z_i$ onto $|-\rangle_i$ with the projections $Q_i^z = \frac{\one_i+\sigma_i^z}{2}$ onto the spin-up configurations $|+\rangle_i$.
\end{rem}
\begin{rem}[\textbf{Translation invariance}]
The result of Theorem \ref{prop: failure of factorization - upper bound} readily generalizes to non-translation invariant models with potentials of the form \begin{equation*}
    \Phi(\{i,j\}) = J_{i,j} P^z_i \sigma^x_j + K_{i,j} \sigma^x_i P^z_j,\qquad |i-j|=1,\,i,j\in\mathbb{Z}^2,
\end{equation*} under suitable hypothesis on the couplings, e.g. $J_{i,j},\,K_{i,j}\geq J > 0$ and $$\sup_{i,j\in\mathbb{Z}^2}\{J_{i,j},K_{i,j}\}<+\infty$$ would be sufficient.    
\end{rem}

\begin{rem}[\textbf{Magnetization bounds}]The bound presented in Thereom \ref{prop: failure of factorization - upper bound} can be equivalently written as an inequality between different macroscopic sizes for the magnetization\eqref{eq: failure of factorization - potential} observable. Indeed, we can write \begin{align*}
    \omega_{X_k(-1,0)}(\one_{\{\frac{1}{2}\}}(m_{X_k(-1,0)}^z)) = &Z_{X_k(-1,0)}^{-1}  
    \leq Z_{X_k}^{-1} e^{2J\beta } \\
    =&\omega_{X_k}(\one_{\{\frac{1}{2}\}}(m_{X_k}^z)) e^{2J\beta}\\
    \leq & \omega_{X_k(-1,0)}(\one_{\{\frac{1}{2}\}}(m_{X_k}^z))e^{2J\beta -C_{J\beta}|X_k| }.
\end{align*} Therefore, we see that changing the magnetization observable only at the boundary produces a macroscopically relevant factor.   \begin{equation*}
    \omega_{X_k(-1,0)}(\one_{\{\frac{1}{2}\}}(m_{X_k}^z)) \geq e^{-2J\beta +C_{J\beta}|X_k| }\omega_{X_k(-1,0)}(\one_{\{\frac{1}{2}\}}(m_{X_k(-1,0)}^z)). 
\end{equation*}
\end{rem}

\begin{rem}[\textbf{Limits of the current approach}] It is not clear at present to what extent one can generalize the above argument to other projections of the magnetization. If we were to consider the projection onto states with one or more spins down, it would be impossible to reproduce the argument used in Theorem \ref{prop: failure of factorization - upper bound}. Indeed, we would have one "activation site" inside the region $X$. This alone would produce a macroscopic contribution to the expectation value in the Gibbs state.
    
\end{rem}

\paragraph{AI discolosure.} Following the release of the first version of this manuscript, Hal Tasaki drew my attention to technical errors in the proofs that had been identified using large language models (LLMs). I subsequently corrected these errors. LLMs were not used in any other part of the manuscript.

\paragraph{Acknowledgements.} I am grateful to Yoshiko Ogata for many illuminating and stimulating discussions on this topic. Without her support, I could not have developed the arguments presented in this manuscript. I also thank Vojkan Jak\v{s}i\'{c} and Anna Szczepanek for drawing my attention to the relevance of Weak Gibbsianity and for their many useful comments. I thank Hal Tasaki for pointing out errors in the first version of this manuscript.

I acknowledge the support of the National Group of Mathematical Physics (GNFM--INdAM).
\paragraph{Data availability statement.}
Data sharing is not applicable to this article as no new data were created or analysed in this study.

\paragraph{Conflict of interest statement.}
The authors certify that they have no affiliations with or involvement in any
organization or entity with any financial interest or non-financial interest in
the subject matter discussed in this manuscript.

\printbibliography

@article{TAPP2026104325,
title = {On the minimum spanning tree distribution in grids},
journal = {European Journal of Combinatorics},
volume = {133},
pages = {104325},
year = {2026},
issn = {0195-6698},
doi = {https://doi.org/10.1016/j.ejc.2025.104325},
url = {https://www.sciencedirect.com/science/article/pii/S0195669825002185},
author = {Kristopher Tapp},
}

@article{Bouch2015,
  author  = {Bouch, Gabriel},
  title   = {Complex-time singularity and locality estimates for quantum lattice systems},
  journal = {Journal of Mathematical Physics},
  volume  = {56},
  number  = {12},
  pages   = {123303},
  year    = {2015},
  doi     = {10.1063/1.4936209}
}

@article{OgataReyBellet2011,
  author  = {Ogata, Yoshiko and Rey-Bellet, Luc},
  title   = {{Ruelle--Lanford} functions and large deviations for asymptotically decoupled quantum systems},
  journal = {Reviews in Mathematical Physics},
  volume  = {23},
  number  = {2},
  pages   = {211--232},
  year    = {2011},
  doi     = {10.1142/S0129055X11004291}
}

@article{JaksicPilletTauber2024Approach,
  author = {Jak{\v{s}}i{\'{c}}, Vojkan and Pillet, Claude-Alain and Tauber, Cl{\'{e}}ment},
  title = {Approach to equilibrium in translation-invariant quantum systems: Some structural results},
  journal = {Annales Henri Poincar{\'{e}}},
  year = {2024},
  volume = {25},
  number = {1},
  pages = {715--749},
  doi = {10.1007/s00023-023-01281-6},
  url = {https://doi.org/10.1007/s00023-023-01281-6},
  eprint = {2204.00440},
  archivePrefix = {arXiv},
}

@misc{JaksicPilletSzczepanek2026ApproachII,
  author = {Jak{\v{s}}i{\'{c}}, Vojkan and Pillet, Claude-Alain and Szczepanek, Anna},
  title = {Approach to equilibrium in translation-invariant quantum systems: Structural results {II}},
  year = {2026},
  howpublished = {arXiv:2610.21475},
  eprint = {2610.21475},
  archivePrefix = {arXiv},
  primaryClass = {math-ph},
  url = {https://arxiv.org/abs/2610.21475},
}

@misc{JaksicPilletSzczepanek2026Conservation,
  author = {Jak{\v{s}}i{\'{c}}, Vojkan and Pillet, Claude-Alain and Szczepanek, Anna},
  title = {Dynamical conservation laws in quantum spin systems},
  year = {2026},
  howpublished = {arXiv:2610.21489},
  eprint = {2610.21489},
  archivePrefix = {arXiv},
  primaryClass = {math-ph},
  url = {https://arxiv.org/abs/2610.21489},
}

@article{JaksicPilletSzczepanek2026Entropy,
  author = {Jak{\v{s}}i{\'{c}}, Vojkan and Pillet, Claude-Alain and Szczepanek, Anna},
  title = {Entropy maximization and weak {Gibbsianity} of quasi-free fermionic states},
  journal = {Letters in Mathematical Physics},
  year = {2026},
  volume = {116},
  number = {4},
  pages = {103},
  doi = {10.1007/s11005-026-02100-9},
  url = {https://doi.org/10.1007/s11005-026-02100-9},
  eprint = {2603.14020},
  archivePrefix = {arXiv},
}

@article{JaksicPilletTauber2024Adiabatic,
  author = {Jak{\v{s}}i{\'{c}}, Vojkan and Pillet, Claude-Alain and Tauber, Cl{\'{e}}ment},
  title = {A note on adiabatic time evolution and quasi-static processes in translation-invariant quantum systems},
  journal = {Annales Henri Poincar{\'{e}}},
  year = {2024},
  volume = {25},
  number = {1},
  pages = {751--771},
  doi = {10.1007/s00023-023-01282-5},
  url = {https://doi.org/10.1007/s00023-023-01282-5},
  eprint = {2204.02177},
  archivePrefix = {arXiv},
}

@article{Araki1969Gibbs,
  author = {Araki, Huzihiro},
  title = {{Gibbs} states of a one dimensional quantum lattice},
  journal = {Communications in Mathematical Physics},
  year = {1969},
  volume = {14},
  number = {2},
  pages = {120--157},
  doi = {10.1007/BF01645134},
  url = {https://doi.org/10.1007/BF01645134},
}

@book{Ruelle1969StatisticalMechanics,
  author = {Ruelle, David},
  title = {Statistical Mechanics: Rigorous Results},
  publisher = {W. A. Benjamin},
  address = {New York},
  year = {1969},
  series = {Mathematical Physics Monograph Series},
  isbn = {0805383603},
}

@article{Yuri2000WeakGibbs,
  author = {Yuri, Michiko},
  title = {Weak {Gibbs} measures for certain non-hyperbolic systems},
  journal = {Ergodic Theory and Dynamical Systems},
  year = {2000},
  volume = {20},
  number = {5},
  pages = {1495--1518},
  doi = {10.1017/S014338570000081X},
  url = {https://doi.org/10.1017/S014338570000081X},
}

@article{Yuri2002WeakGibbs,
  author = {Yuri, Michiko},
  title = {Weak {Gibbs} measures and the local product structure},
  journal = {Ergodic Theory and Dynamical Systems},
  year = {2002},
  volume = {22},
  number = {6},
  pages = {1933--1955},
  doi = {10.1017/S0143385702000780},
  url = {https://doi.org/10.1017/S0143385702000780},
}

@article{PfisterSullivan2018WeakGibbs,
  author = {Pfister, Charles-Edouard and Sullivan, Wayne G.},
  title = {Weak {Gibbs} measures and large deviations},
  journal = {Nonlinearity},
  year = {2018},
  volume = {31},
  number = {1},
  pages = {49--53},
  doi = {10.1088/1361-6544/aa99a3},
  url = {https://doi.org/10.1088/1361-6544/aa99a3},
  eprint = {1705.06939},
  archivePrefix = {arXiv},
}

@article{PfisterSullivan2020Asymptotic,
  author = {Pfister, Charles-Edouard and Sullivan, Wayne G.},
  title = {Asymptotic decoupling and weak {Gibbs} measures for finite alphabet shift spaces},
  journal = {Nonlinearity},
  year = {2020},
  volume = {33},
  number = {9},
  pages = {4799--4817},
  doi = {10.1088/1361-6544/ab8fb7},
  url = {https://doi.org/10.1088/1361-6544/ab8fb7},
}

@book{BratteliRobinson1987Vol1,
  author = {Bratteli, Ola and Robinson, Derek W.},
  title = {Operator Algebras and Quantum Statistical Mechanics 1: {$C^*$}- and {$W^*$}-Algebras. Symmetry Groups. Decomposition of States},
  edition = {2},
  publisher = {Springer},
  address = {Berlin, Heidelberg},
  year = {1987},
  isbn = {978-3-540-17093-8},
  doi = {10.1007/978-3-662-02520-8},
  url = {https://doi.org/10.1007/978-3-662-02520-8},
}

@book{BratteliRobinson1997Vol2,
  author = {Bratteli, Ola and Robinson, Derek W.},
  title = {Operator Algebras and Quantum Statistical Mechanics 2: Equilibrium States. Models in Quantum Statistical Mechanics},
  edition = {2},
  publisher = {Springer},
  address = {Berlin, Heidelberg},
  year = {1997},
  isbn = {978-3-540-61443-2},
  doi = {10.1007/978-3-662-03444-6},
  url = {https://doi.org/10.1007/978-3-662-03444-6},
}

@article{ArakiIon1974Equivalence,
  author  = {Araki, Huzihiro and Ion, P. D. F.},
  title   = {On the equivalence of {KMS} and {Gibbs} conditions for states of quantum lattice systems},
  journal = {Communications in Mathematical Physics},
  year    = {1974},
  volume  = {35},
  number  = {1},
  pages   = {1--12},
  doi     = {10.1007/BF01646450},
  url     = {https://doi.org/10.1007/BF01646450}
}

@incollection{Araki1976PositiveCone,
  author    = {Araki, Huzihiro},
  title     = {Positive cone, {Radon--Nikod\'ym} theorems, relative {Hamiltonian} and the {Gibbs} condition in statistical mechanics. {An} application of the {Tomita--Takesaki} theory},
  editor    = {Kastler, Daniel},
  booktitle = {{$C^*$}-Algebras and Their Applications to Statistical Mechanics and Quantum Field Theory},
  series    = {Proceedings of the International School of Physics ``Enrico Fermi''},
  volume    = {60},
  publisher = {North-Holland},
  address   = {Amsterdam},
  year      = {1976},
  pages     = {64--100},
  note      = {Course LX, Varenna, 1973}
}

@misc{Tasaki2024Bouch,
  author        = {Tasaki, Hal},
  title         = {Trees that can be grown in ``too many'' ways:
                   {Bouch}'s construction and a bound for every size},
  year          = {2026},
  eprint        = {2412.16912},
  archivePrefix = {arXiv},
  primaryClass  = {math-ph},
  note          = {Version 4. Originally posted in December 2024},
  doi           = {10.48550/arXiv.2412.16912},
  url           = {https://arxiv.org/abs/2412.16912v4}
}

@article{ShiraishiTasaki2026,
  author        = {Shiraishi, Naoto and Tasaki, Hal},
  title         = {The {$S=\frac{1}{2}$} {XY} and {XYZ} Models
                   on the Two- or Higher-Dimensional Hypercubic Lattice
                   Do Not Possess Nontrivial Local Conserved Quantities},
  journal       = {Annales Henri Poincar\'e},
  year          = {2026},
  doi           = {10.1007/s00023-026-01699-8},
  eprint        = {2412.18504},
  archivePrefix = {arXiv},
  primaryClass  = {cond-mat.stat-mech}
}

\end{document}